%% file: main.tex
\documentclass[letterpaper,twocolumn,10pt]{article}
\usepackage{usenix}

\usepackage[underline=false]{pgf-umlsd}
\usepackage{msc}
\usepackage{xspace}
\usepackage{textcomp}
\usepackage{tabularx}
\usepackage{graphicx}
\usepackage{booktabs}
\usepackage{circledsteps}
\usepackage{enumitem}
\setitemize{noitemsep,topsep=0pt,parsep=0pt,partopsep=0pt}
\setlist{nolistsep}
\usepackage[ruled]{algorithm2e}
\usepackage{amsmath}
\usepackage{colortbl}
\usepackage{multirow}
\usepackage{pifont}
\usepackage{bm}
\usepackage{hyperref}
\usepackage{appendix}
\usetikzlibrary{positioning, shapes.geometric, arrows.meta}

\hypersetup{
    colorlinks=true,
    linkcolor=blue,
    urlcolor=blue,
    citecolor=blue,
    }

\def \showcomments {Show comments}
\ifdefined\showcomments
\usepackage{todonotes}
\usepackage{letltxmacro}
\LetLtxMacro{\todonote}{\todo}
\renewcommand{\todo}[2][]
{\todonote[inline, caption={#2}, size=\footnotesize, #1]
{\renewcommand{\baselinestretch}{0.5}\selectfont#2\par}}
\else
\usepackage[disable]{todonotes}
\fi

\ifdefined\showchanges
\newcommand{\changed}[1]{\color{red}{#1}\color{black}\xspace}
\else
\newcommand{\changed}[1]{#1\xspace}
\fi

\newcommand{\taggedpara}[1]{\vspace{1mm}\noindent\textbf{#1.}}
\newcommand{\utaggedpara}[1]{\vspace{1mm}\noindent\underline{#1.}}

\newcommand{\sys}{\texttt{Gyokuro}\xspace}

\begin{document}

\input{notations}

\date{}

\title{\Large \bf \sys: Source-assisted Private Membership Testing using \\ Trusted Execution Environments}

\author{
{\rm Yoshimichi Nakatsuka}\\
ETH Zurich
\and
{\rm Nicolas Dutly}\\
ETH Zurich
\and
{\rm Kari Kostiainen}\\
ETH Zurich
\and
{\rm Srdjan \v Capkun}\\
ETH Zurich
}

\maketitle

\input{contents/00-abstract}
\input{contents/01-intro}
\input{contents/02-background}

\input{contents/03a-problem-statement}
\input{contents/04a-overview}
\input{contents/04b-design}

\input{contents/06-evaluation}

\input{contents/07-application}
\input{contents/10-conclusion}

\cleardoublepage
\appendix
\section*{Ethical Considerations}
The implementation and evaluation of this work does not present ethical issues as it did not involve human subjects or measurements that involve in-production services.
Our stakeholder-based risk analysis of the entities involved in \sys is as follows.
\begin{itemize}
    \item Data Source: As the entity that generates data and submits them to the server, the use of \sys does not present any privacy or security risks to the data source.
    \item Server: There are no risks for the entities that interact with the server. Entities involved in law enforcement agencies may have the risk of being eluded due to the lack of information on the server-side; however, client interests remain present on the client-side which law enforcement agencies can investigate via a valid search warrant. Therefore, we do not consider this to be a significant risk.
    \item Database Operator: The database operator experiences no risks as it simply stores data items and responds to requests from the monitor.
    \item Client: There are no risks for the client in using \sys as their private data is not exposed to any external party.
    \item Monitor: The monitor does not encounter any risks as its activities do not differ from those of an existing monitor.
\end{itemize}

\section*{Open Science}
The following artifacts are published at \cite{AnonymousGithubOnline}:
\begin{itemize}
    \item The complete server-side logic of \sys for Data submission and Storage \& Processing, as well as Membership testing \& Consistency checking protocols which were used to produce the results shown in Section~\ref{sec:eval}.
    \item A sample database implementation that interacts with the server-side logic to store submitted data.
    \item Client-side scripts used to conduct the evaluation.
    \item Other pieces of data (e.g., server keys) necessary for the server-side to operate.
\end{itemize}

{
    \footnotesize
    \bibliographystyle{plain}
    \bibliography{references}
}

\normalsize

\input{contents/12-appendix}

\end{document}

%% file: notations.tex
\newcommand{\dds}{d}
\newcommand{\cntpor}{cnt_{POR}}
\newcommand{\cntpop}{cnt_{POP}}
\newcommand{\sks}{sk_{T}}
\newcommand{\pks}{pk_{T}}
\newcommand{\skds}{sk_{DS}}
\newcommand{\pkds}{pk_{DS}}
\newcommand{\skm}{sk_{M}}
\newcommand{\pkm}{pk_{M}}

%% file: contents/00-abstract.tex
\begin{abstract}

Private Membership Testing (PMT) protocols enable clients to verify whether a certain data item is included in a database without revealing the item to the database operator or other external parties.
This paper examines \emph{Source-assisted} PMT (SPMT), in which clients leverage compact data source-provided information issued when the data item is first submitted to the database.
SPMT is relevant in applications such as certificate transparency and supply-chain auditing; yet, designing an approach that is efficient, scalable, and privacy-preserving remains a challenge.

This work presents \sys, which takes a different approach to conventional membership testing schemes.
Instead of requesting the server to produce a proof attesting that a certain data item exists in the database, we leverage Trusted Execution Environments (TEEs) to produce proofs demonstrating that the server has made enough progress to add the data item to the database. With the help of existing monitoring services, clients can infer that no items have been removed from the database.
This allows \sys to provide strong privacy guaranties and achieve high efficiency, as a client's membership testing query does not include any information regarding their interests, and eliminates the need for complex and inefficient protection mechanisms.
Additionally, this approach enables membership testing on large-scale databases, since the communication and computation required are independent of the database size.
Our evaluations show practical feasibility, achieving 7 ms membership testing latency and throughput of around 1400 requests/sec/core.

\end{abstract}

%% file: contents/01-intro.tex
\section{Introduction} 
\label{sec:intro}

Verifying whether a certain piece of data is included in a database with the aid of an external party (i.e., monitor) while protecting the verifier's (i.e., client's) privacy from the prover (i.e., server) is called ``Private Membership Testing'' (PMT)~\cite{meskanen2015bloom,tamrakar2017circle}.
    Many prior work has proposed PMT protocols that leverage techniques such as Private Information Retrieval (PIR)~\cite{chor1998pir,chor1998keywords,kushilevitz1997replication,gentry2005single,ishai_batch_2004,rawat_batch_2016,b_paterson_combinatorial_2009,angel_pir_2018,lueks_sublinear_2015,angel_unobservable_2016,mughees_vectorized_2023,boyle_can_2017,nguyen_client-efficient_2025,lazzaretti_near-optimal_2023,zhou_optimal_2023,zhou_piano_2024,kogan_private_2021,corrigan-gibbs_private_2020,shi_puncturable_2021,beimel_reducing_2000,corrigan-gibbs_single-server_2022,canetti_towards_2017,lazzaretti_treepir_2023}, Private Set Intersection (PSI)~\cite{chen_fast_2017,al_badawi_implementation_2021,chen_labeled_2018,ren_privacy-preserving_2021,caudhari_securing_2021,raghuraman_blazing_2022,kolesnikov_efficient_2016,hemenway_falk_private_2019,pinkas_psi_2020,rindal_vole-psi_2021}, Oblivious RAM (ORAM)~\cite{backes2012obliviad,vadapalli2023duoram,falk2022doram,jiang2021oram}, homomorphic encryption~\cite{ramezanian2020low,chielle2021realtime}, and trusted computing~\cite{tamrakar2017circle,wang2006trustedpir,iliev2005trustedpir}.

However, designing a PMT protocol that is efficient, scalable, and privacy-preserving remains a challenge.
    In this work, we focus on a sub-variant of the PMT protocol, which we denote as \emph{Source-assisted} PMT (SPMT).
    The unique aspect of this protocol is that clients utilize information they received from the data source, denoted as \emph{source-assisted information}, in addition to the data item during membership testing.
SPMT protocols are useful in various applications, such as certificate transparency (CT) log auditing, supply-chain auditing, data provenance verification, and public document auditing.
    For example, in the context of drug supply-chain auditing, consumers (\textit{clients}) check whether the supply-chain information (\textit{source-assisted information}) of the drug they purchased is issued by valid pharmaceutical supply-chain partners (\textit{data sources}) and is stored in valid public logs (\textit{server/database}) without revealing the drug label (\textit{private data item}).
    Supply-chain information could, e.g., be presented to consumers in the form of a QR code found on the medicine package.
    Consumers would then engage in an SPMT protocol with the public log using this QR code to verify that the public log has stored the supply-chain information correctly, while receiving assistance from regulatory agencies (\textit{monitors}) that inspect whether supply-chain partners comply with regulations.

Assisting clients with source-assisted information does not make the SPMT problem trivial, as designing an SPMT protocol that balances efficiency, scalability, and privacy remains a challenging task.
The complexity of the SPMT problem is also illustrated by the lack of a known method to privately audit CT logs that is efficient and scalable~\cite{meiklejohn_sok_2022}.
Current browser implementations either forego CT log auditing altogether~\cite{keeler_firefox_2025} or perform it in a semi-private manner, providing only $k$-set anonymity~\cite{deblasio_how_2023}.

\taggedpara{Alternative approach \& its issues}
    If we can design a system that does not require clients to interact with any third party by leveraging the source-assisted information, then this system is, by design, privacy-preserving.
A method that has been overlooked by related efforts is leveraging \emph{the passage of time}; that is, if a server can guarantee that it will store data items before a specific deadline, then the client can locally check whether enough time has passed during membership testing.
    This requires the server to guarantee \emph{execution integrity} (i.e., it will follow the protocol) and \emph{execution timeliness} (i.e., it will finish executing before a certain deadline).
However, upholding the two guarantees under actively malicious servers, which is assumed by the SPMT protocol, is no easy task.
    Hardware-based confidential computing technologies (e.g., Trusted Execution Environments, TEEs) have demonstrated that guaranteeing execution integrity under malicious environments is possible; however, ensuring execution timeliness under such conditions is, to the best of our knowledge, impossible, even with the aid of TEEs~\cite{alder2023about}.

\taggedpara{Our solution}
Motivated by the above discussion, we propose \sys, a system that takes a different approach.
    As currently available TEEs cannot uphold timeliness guarantees, we take the viable alternative which is to check whether the TEE is making \emph{sufficient progress}.
    There are several ways to realize this in practice, and this work uses an approach based on monotonically increasing counters stored within the TEE.

\changed{
\sys is composed of two protocols, \textit{Upload} and \textit{Testing}.
For every data item generated by the data source, \sys runs the Upload protocol, shown below.
    When the TEE receives a data item from the data source, it generates a digital signature over the data item and the current counter value, denoted as ``Proof Of Reception'' ($POR$).
    Once the TEE receives a fixed number of data items, it stores them to an external database and increments the counter.
        The TEE also updates an accumulator data structure, which compresses the entire history of data items submitted to the TEE into a single structure serving as the master record of all submitted data items.

For the data item which the client wants to test, \sys runs the Testing protocol, shown below.
    The client receives one of the data item sent by the data source and its accompanying $POR$ (source-assisted information).
    The monitor periodically downloads the entire database and updates its accumulator.
    During membership testing, the client requests the TEE for a digital signature over the current counter value, denoted as ``Proof Of Processing'' ($POP$).
        The client verifies $POR$ and $POP$ using the TEE's public key, and checks whether the counter value in $POP$ is greater than in $POR$.
        If it is, then the client concludes that the data item has been processed by the TEE and sent to the database for storage.
    The client requests accumulator values from the TEE and monitor. If the TEE's and monitors' accumulator values match, the client can safely conclude that the tested data item is present in the database.
}

\taggedpara{Benefits and limitations}
One \emph{novel aspect} of this work is that the client does not send the data item or any of its derivatives to the database operator during membership testing.
    Instead, the client infers that the data item is included in the database by comparing the counter values and accumulator values it received from the TEE and monitors.
    This is conceptually different from every prior PMT approach.

As with previous work, ensuring \emph{client privacy} is our utmost priority.
    In our design, the membership testing query is independent of the client's interest.
    This means that even if the TEE is targeted with side-channel attacks, the queried data item cannot leak to the untrusted server, as it is never sent to the TEE.
    This is a significant advantage, given the increase in side-channel attacks targeting TEE confidentiality~\cite{liu2015last,xu2015controlled,wang2017leaky,chen2021voltpillager,brasser2017software,dall2018cachequote,gotzfried2017cache,moghimi2017cachezoom,kocher2019spectre,moritz2018meltdown,bluck2018foreshadow,murdock2020plundervolt}.

Another benefit of \sys is its \emph{efficiency}.
    To hide database access patterns from revealing the client query, TEE-based PMT solutions typically rely on expensive cryptographic primitives, such as ORAM~\cite{wang2006trustedpir,iliev2005trustedpir}, or simply touch every database record~\cite{tamrakar2017circle}, which puts a toll on the solutions' performance.
    Our design does not have this shortcoming because the TEE does not communicate with the database during membership testing at all, allowing for extremely fast query processing, which we show is around 7 ms for both Upload and Testing protocols in our experimental setup.

\emph{Scalability} is also an important feature of \sys, which we show to be 1021 and 1432 requests/sec/core for Upload and Testing protocols, respectively.
    This is because the amount of communication and computation required in the proposed system is independent of the database size.
    This allows \sys to support large-scale systems, e.g., CT logs storing billions of certificates.

Despite such numerous benefits, \sys has one notable limitation.
    As our solution relies on TEEs, there is no guarantee that the required execution integrity property will not be compromised in the future.
    Purely cryptographic PMT solutions are free from this limitation.

\taggedpara{Contributions}
 In summary, the main contributions of this work are as follows.

\vspace{1mm}
\begin{itemize}
    \item \textbf{Problem identification.} We observe that the SPMT problem has not been widely studied but appears in popular applications such as certificate transparency and supply-chain auditing (Section~\ref{sec:problemstatement}).
    \item \textbf{Novel solution.} We propose \sys, a novel TEE-based SPMT scheme that does not require the client to disclose their private information to the database operator, TEE, or any third party (Sections~\ref{sec:designoverview} and \ref{sec:designdetails}).
    \item \textbf{Performance evaluation.} We demonstrate low data submission and membership test latencies, on the order of several milliseconds, and high system throughput by serving thousands of requests per second per CPU core (Section~\ref{sec:eval}).
    \item \textbf{Security analysis}. Our analysis shows that \sys guarantees membership testing correctness and strong client privacy protection (Section~\ref{sec:eval:security}).
\end{itemize}

\vspace{1mm}
Our open-source implementation is available at \cite{AnonymousGithubOnline}.

%% file: contents/02-background.tex
\section{Background and Related Work} \label{sec:background}

\subsection{Private Membership Testing}
Private Membership Testing (PMT) is a protocol that allows clients to verify whether a specific data item exists in a database without disclosing their interest in that item to any external entity.
Here, we highlight primitives proposed in prior work that can be used to protect client privacy during membership testing.
A comparison of these primitives with our solution \sys is presented in Table~\ref{tab:related-work}, while Table~\ref{tab:big-o} shows comparison of the computational and communication complexity with respect to the size of the database.

\begin{table*}[th]
\centering
\footnotesize
\caption{Comparison with related work. ``Low'' computation/communication indicates constant cost and ``High'' scalability indicates low online computation and communication.} \label{tab:related-work}
\setlength{\extrarowheight}{2pt} %

\newcolumntype{Y}{>{\centering\arraybackslash}X} %
\newcolumntype{Z}{>{\centering\arraybackslash}m{85pt}} %

\makeatletter
\newcommand*\rot[2][66pt]{%
  \rotatebox{90}{%
    \begin{tabular}[c]{@{}p{#1}@{}}%
      \raggedright #2\strut
    \end{tabular}%
  }%
}
\makeatother

\begin{tabularx}{\textwidth}{cZYYYYYYYYYY}
\toprule
& & \multicolumn{2}{c}{Trust assumptions} & \multicolumn{6}{c}{Performance} & \multicolumn{2}{c}{Security \& Privacy} \\
\cmidrule(lr){3-4} \cmidrule(lr){5-10} \cmidrule(lr){11-12}
& & \rot{No non-colluding servers required} & \rot{No trusted computing required} & \rot{Low offline computation cost} & \rot{Low online computation cost} & \rot{Low communication cost} & \rot{High scalability} & \rot{No client-side storage required} & \rot{No database pre- processing required} & \rot{Resistant to side-channel attacks} & \rot{Privacy-preserving} \\ 
\midrule
\rowcolor[HTML]{EFEFEF}
\cellcolor[HTML]{FFFFFF} & Vanilla PIR~\cite{chor1998pir} & \ding{55} & \ding{51} & \ding{51} & \ding{55} & \ding{55} & \ding{55} & \ding{51} & \ding{51} & \ding{51} & \ding{51} \\
& Single-server PIR~\cite{gentry2005single} & \ding{51} & \ding{51} & \ding{55} & \ding{55} & \ding{55} & \ding{55} & \ding{51} & \ding{55} & \ding{51} & \ding{51} \\
\rowcolor[HTML]{EFEFEF}
\cellcolor[HTML]{FFFFFF} & Batched PIR~\cite{mughees2023vectorized} & \ding{51} & \ding{51} & \ding{55} & \ding{55} & \ding{51} & \ding{55} & \ding{51} & \ding{55} & \ding{51} & \ding{51} \\
& Online-Offline PIR~\cite{nguyen_client-efficient_2025} & \ding{51} & \ding{51} & \ding{55} & \ding{55} & \ding{55} & \ding{55} & \ding{55} & \ding{55} & \ding{51} & \ding{51} \\
\rowcolor[HTML]{EFEFEF}
\cellcolor[HTML]{FFFFFF} & PSI~\cite{raghuraman_blazing_2022} & \ding{51} & \ding{51} & \ding{51} & \ding{55} & \ding{55} & \ding{55} & \ding{51} & \ding{51} & \ding{51} & \ding{51} \\
& ORAM~\cite{jiang2021oram} & \ding{51} & \ding{51} & \ding{55} & \ding{55} & \ding{55} & \ding{55} & \ding{51} & \ding{55} & \ding{51} & \ding{51} \\
\rowcolor[HTML]{EFEFEF}
\cellcolor[HTML]{FFFFFF} & HE~\cite{chielle2021realtime} & \ding{51} & \ding{51} & \ding{55} & \ding{51} & \ding{51} & \ding{51} & \ding{51} & \ding{55} & \ding{51} & \ding{51} \\
\multirow{-8}{*}{PMT} & Trusted computing~\cite{tamrakar2017circle} & \ding{51} & \ding{55} & \ding{55} & \ding{55} & \ding{51} & \ding{55} & \ding{51} & \ding{55} & \ding{55} & \ding{55} \\ 
\midrule
\rowcolor[HTML]{EFEFEF}
\cellcolor[HTML]{FFFFFF} & Proof of inclusion~\cite{meiklejohn_sok_2022} & \ding{51} & \ding{51} & \ding{51} & \ding{55} & \ding{55} & \ding{55} & \ding{51} & \ding{51} & \ding{51} & \ding{55} \\
\multirow{-2}{*}{SPMT} & \textbf{This work} & \ding{51} & \ding{55} & \ding{51} & \ding{51} & \ding{51} & \ding{51} & \ding{51} & \ding{51} & \ding{51} & \ding{51} \\ 
\bottomrule
\end{tabularx}

\end{table*}

\begin{table}[th]
\centering
\footnotesize
\caption{Per-query computation and communication complexity with respect to database size $M$} \label{tab:big-o}

\newcolumntype{Y}{>{\centering\arraybackslash}X} %
\newcolumntype{Z}{>{\centering\arraybackslash}m{65pt}} %

\begin{tabularx}{\columnwidth}{ZYYYY}
\toprule
                                            & Offline Computation & Online Computation & Communi-cation \\
\midrule                          
\rowcolor[HTML]{EFEFEF}
Vanilla PIR~\cite{chor1998pir}     & N/A & $O(M)$ & $O(\sqrt{M})$ \\
Single-serv. PIR~\cite{gentry2005single}     & $O(M)$ & $O(\sqrt{M})$ & $O(log~M)$ \\
\rowcolor[HTML]{EFEFEF}
Batched PIR~\cite{mughees2023vectorized}     & $O(M)$ & $O(M)$ & $O(1)$ \\
OO-PIR~\cite{nguyen_client-efficient_2025}     & $O(M)$ & $O(\sqrt{M})$ & $O(\sqrt{M})$ \\
\rowcolor[HTML]{EFEFEF}
PSI~\cite{raghuraman_blazing_2022}          & N/A & $O(M)$ & $O(M~log~M)$ \\
ORAM~\cite{jiang2021oram}                   & $O(M) $& $O(log~M)$ & $O(M~log~M)$ \\
\rowcolor[HTML]{EFEFEF}
HE~\cite{chielle2021realtime}               & $O(M^{2}~log~M)$ & $O(1)$ & $O(1)$ \\
Trusted comp.~\cite{tamrakar2017circle} & $O(M)$ & $O(M)$ & $O(1)$ \\ %
\midrule
\rowcolor[HTML]{EFEFEF}
Proof of inc.~\cite{meiklejohn_sok_2022} & N/A & $O(log~M)$ & $O(log~M)$ \\
\textbf{This work}                          & \textbf{N/A} & $\bm{O(1)}$ & $\bm{O(1)}$ \\
\bottomrule
\end{tabularx}
\end{table}

\taggedpara{PIR}
Private Information Retrieval (PIR) enables clients to \emph{retrieve} specific items from a database without revealing their interest. Privacy-preserving retrieval schemes could also be used for private \emph{membership testing}, although the intended usage of such schemes is slightly different.
    
    Early PIR schemes~\cite{chor1998pir,chor1998keywords} depended on two synchronized (i.e., communicating) yet non-colluding databases, which can be difficult to realize in practice.
    This led to the introduction of single-server PIR schemes~\cite{kushilevitz1997replication,gentry2005single}, which, however, incurred a high communication cost.
    Recent schemes, such as Batched PIR~\cite{ishai_batch_2004,rawat_batch_2016,b_paterson_combinatorial_2009,angel_pir_2018,lueks_sublinear_2015,angel_unobservable_2016,mughees_vectorized_2023} and Online-Offline (OO) PIR~\cite{boyle_can_2017,nguyen_client-efficient_2025,lazzaretti_near-optimal_2023,zhou_optimal_2023,zhou_piano_2024,kogan_private_2021,corrigan-gibbs_private_2020,shi_puncturable_2021,beimel_reducing_2000,corrigan-gibbs_single-server_2022,canetti_towards_2017,lazzaretti_treepir_2023}, improve performance, although there is a trade-off between communication cost and server- and client-side computation, with additional compromises in areas such as client storage and offline computation for OO PIR schemes.

\taggedpara{PSI}
Private Set Intersection (PSI) is a specific type of secure multi-party computation, where two or more parties provide their private sets as inputs and receive their intersection as outputs, but nothing else.
    In the context of PMT, the client engages in a PSI protocol with the server by presenting a set with a single item, while the server provides the content of the database as its set.

    The main drawback of PSI-based approaches is their limitations in performance due to the cryptographic primitives they leverage.
    For instance, HE- and Pairing-based PSI~\cite{chen_fast_2017,al_badawi_implementation_2021,chen_labeled_2018,ren_privacy-preserving_2021,caudhari_securing_2021} trade off communication costs for computation costs, while OT-, OPRF-, and OKVS-based PSI~\cite{raghuraman_blazing_2022,kolesnikov_efficient_2016,hemenway_falk_private_2019,pinkas_psi_2020,rindal_vole-psi_2021} vise versa.
    Moreover, computational costs increase significantly in fully malicious settings, as protocols must introduce additional protection measures, including zero-knowledge proofs or oblivious key-value stores.
    For a complete overview of the PSI work, we refer to \cite{morales2023psi}.

\taggedpara{ORAM}
Oblivious RAM (ORAM) hides access patterns within the data structure, and its performance has been shown to be relatively good.
    Using this property, ORAM can hide PMT queries when accessing the database content.
    However, ORAM-based PMT schemes are limited in their scalability, as they do not natively support multiple clients~\cite{backes2012obliviad,jiang2021oram}.
    Several recent works have proposed multi-client ORAMs~\cite{vadapalli2023duoram,falk2022doram}, but they are still limited to a small number of clients.

\taggedpara{HE}
PMT methods that leverage Homomorphic Encryption (HE)~\cite{chielle2021realtime,ramezanian2020low} encrypt the client query, run the encrypted query over the database content to produce an encrypted response, and then send the encrypted response to the client, which decrypts it to retrieve the PMT result.
    Since client queries are evaluated in their encrypted form, no information is leaked to the server.
    However, HE-based approaches are impractical for large databases due to the high offline computation cost as well as the insertion/deletion latencies.
    Furthermore, these schemes require database preprocessing to support encrypted queries, making them less suitable for applications with frequently updated databases.

\taggedpara{Trusted computing}
Previous works have leveraged server-side trusted computing to solve the PMT problem~\cite{tamrakar2017circle,wang2006trustedpir,iliev2005trustedpir}. In these approaches, clients encrypt queries such that they can only be decrypted within a trusted computing platform, e.g., a TEE.
    However, this causes PMT queries to exist in plaintext within the TEE, rendering them vulnerable to side-channel attacks.
    Moreover, to prevent database access patterns from leaking information about the query, these approaches either rely on privacy-preserving protocols and data structures or necessitate the traversal of the entire database, with the former compromising scalability and the latter, performance.

\taggedpara{Out of scope: Anonymous networks}
Several PMT protocols leverage anonymous networks, such as Tor and mix-nets, to allow clients to hide their identity when submitting their PMT query to the server~\cite{dahlberg_privacy-preserving_2021,eskandarian_certificate_2017,yadav_automatic_2022}.
    However, we do not consider this approach ``private'', since the query is not hidden from the server.
    With knowledge of client queries, the server can still estimate the overall distribution of the database access patterns and launch correlation attacks. 
    Therefore, we consider anonymous network-based approaches orthogonal to our work.

\subsection{Monitors} \label{sec:background:monitors}
This work utilizes existing third-party monitors. 
We point out that any PMT scheme that seeks to remain secure against \emph{split-view attacks}, where the malicious server creates multiple versions of the database and presents different views to different parties, requires support from some external entity (e.g., monitors~\cite{laurie_certificate_2021}, gossipping between clients~\cite{laurent2015gossip}, blockchains~\cite{catena2017alin,ethiks2016bonneau}) that enables the client to verify that it is performing the membership test protocol with the correct database version.

Let us take Certificate Transparency (CT) as an example. 
    Assume that a malicious CT log operator keeps two copies of the log, A and B, with log A containing a maliciously issued certificate and log B not.
    The log operator could show log A to clients while presenting log B to monitors in an attempt to convince clients that the malicious certificate is logged for public scrutiny while preventing monitors from detecting the certificate.
    To identify this type of attack, clients and monitors must share their views of the database.
    Such attacks are not exclusive to our work; indeed, all previous works discussed in Section~\ref{sec:background} must rely on an external entity to detect this attack.

Monitors that check database consistency are already available in existing ecosystems, such as CT.
    In addition to allowing domain owners to check for misissued certificates, the CT specification enables monitors to maintain copies of entire logs~\cite{laurie_certificate_2014,laurie_certificate_2021}, a practice adopted by many CT monitors~\cite{censys,crtsh,facebookmonitor,certspotter,merklemap}.
    During this process, the specification requires them to calculate an accumulator using the log copy to ensure that the CT log operator does not remove any certificates.
        In the case of CT, this accumulator is an MHT root, which is signed and periodically published by the CT log (called Signed Tree Head, STH).
        To verify that no data items are removed from the database, a monitor downloads every data item from the database, reconstructs the MHT root, and compares it with the STH.
    Although our solution \sys leverages hash chains, it can equally leverage MHT roots.

\subsection{Trusted Execution Environments} \label{sec:background:tees}
TEEs protect security-sensitive code and data from untrusted entities.
Examples of TEEs include Intel SGX~\cite{anati2013innovative}, Intel TDX~\cite{TDX}, AMD SEV~\cite{SEV,SEV-SNP}, ARM TrustZone~\cite{TrustZone}, and ARM CCA~\cite{CCA}. The following functionalities are typically provided.

\utaggedpara{Isolated execution}
TEEs are isolated from all other software, including privileged ones, such as the OS and hypervisor.
Most prominently, data within the TEE can only be accessed by the code running inside the TEE (\emph{confidentiality}).
Moreover, the code inside the TEE provides well-defined entry points, preventing adversaries from running arbitrary parts of the code (\emph{execution integrity}).
To mitigate the wide range of side-channel attacks proposed in the literature~\cite{liu2015last,xu2015controlled,wang2017leaky,chen2021voltpillager,brasser2017software,dall2018cachequote,gotzfried2017cache,moghimi2017cachezoom,kocher2019spectre,moritz2018meltdown,bluck2018foreshadow,murdock2020plundervolt}, it is commonly accepted that constant-time coding techniques provide some form of protection.
Although applying constant-time coding in cryptographic libraries has seen immense progress, applying this to \emph{general-purpose} programs is shown to be difficult~\cite{zhang2025farfetchdsidechannelanalysisframework} and error-prone~\cite{jancar_2022_constant,marcel_2024_constant}, and therefore relying on TEEs to protect the confidentiality of arbitrary data is no longer ideal.
The majority of attacks against TEEs target its confidentiality aspect, while there has been little focus on attacking its integrity.

\utaggedpara{Remote attestation}
Remote attestation enables TEEs to generate attestation proofs, providing remote parties with strong assurances about the TEE and the code running within it.
Specifically, the TEE: (1) demonstrates that it is a genuine TEE running on genuine hardware, and (2) describes the code in a cryptographically verifiable manner.
Using the proof, the remote party decides whether to trust the TEE and its outputs.

%% file: contents/03a-problem-statement.tex
\section{Problem Statement} \label{sec:problemstatement}

A significant challenge in designing a PMT scheme is to balance efficiency, scalability, and client privacy.
This work focuses on a specific variant of the PMT protocol, referred to as \emph{Source-assisted} PMT (SPMT), where clients engage in the protocol with information provided by the data source in addition to the issued data item.
    Although SPMT protocols have not received much attention from the research community, they can be utilized in various applications, including CT log auditing, supply chain auditing, data provenance verification, and public document auditing.
    We discuss such applications and their suitability to our solution in Section~\ref{sec:applications}.

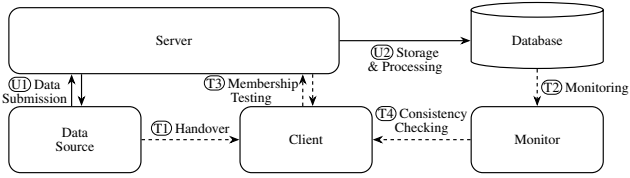
\begin{figure}[t]
    \centering
    \resizebox{\columnwidth}{!}{\input{figures/spmt}}
    \caption{SPMT system model and main operations. \changed{Solid arrows represent Upload protocols and dashed arrows represent Testing protocols.}}
    \label{fig:pmt}
\end{figure}

\subsection{System and Threat Model}

The SPMT protocol involves the following five entities: data source, client, server, database, and monitor, as illustrated in Figure~\ref{fig:pmt}.
We assume that clients are trusted.
Monitors are considered honest-but-curious, i.e., they will faithfully provide their view of the database to clients, but may attempt to violate their privacy.
We consider servers, database hosts, and data sources fully malicious.
The goal of the malicious server and the database host is to selectively remove a certain piece of data from being processed and stored without being caught.
    For instance, they may remove the target data from the database \emph{before} the monitor downloads it so that the existence of the target data is not known to any external auditors.
Additionally, the goal of the malicious server is to violate the client's privacy.
The goal of a malicious data source is to deceive clients into believing that a specific piece of data is stored in the database, when in fact it is not.
We assume that data sources and monitors safeguard their signing keys to prevent impersonation and thus will not be leaked.
Finally, we assume that data generated by data sources and that the database is public.

\utaggedpara{TEE Trust Model}
Although general SPMT solutions do not necessarily need to use TEEs, the proposed approach does; therefore, we define the TEE trust model as follows.
We assume that TEEs are trusted for their execution integrity and secure state.
    The TEE host (e.g., an untrusted OS) has full control over the network, scheduling, and memory management; therefore, it can arbitrarily delay network packets, temporarily halt TEEs, and learn TEE memory access patterns.
    Providing TEE state continuity~\cite{strackx2016ariadne}, rollback attack prevention~\cite{chu2025rollbaccine,matetic2017rote,angel_nimble_2023,hartono_crisp_2024}, and intra-machine fork detection~\cite{briongos2023forking,strackx2016ariadne,parno_memoir_2011}, as well as inter-machine fork detection~\cite{matetic2017rote,niu_narrator_2022}, are considered out of scope but are assumed to be integrated into the TEE.
    Additionally, all cryptographic primitives used within the TEE are assumed to implement constant-time coding techniques and therefore do not leak keys when used. Any other non-cryptographic computations performed by the TEE may leak processed data.
Denial-of-service attacks against TEEs and attacks that modify TEE execution paths (e.g., return-oriented programming attacks) are considered out of scope.

\subsection{Main Operations}
\changed{
The SPMT protocol consists of the following two protocols: (1) Upload and (2) Testing.
Upload protocols are \emph{standard operations}, meaning that these protocol run for every data item submitted by a data source.
Meanwhile, Testing protocols are \emph{conditional operations}, where the condition is triggered if the client obtains a data item it is interested in (e.g., client purchases a drug).
    This means that Testing protocols run for limited selection of data items under which the client requested from the data source.
The two protocols can be further broken down into the following sub-protocols:
}

\begin{enumerate}[start=1,label={\Circled{U\arabic*}}]
    \item \textit{Data submission:}
        The data source generates a data item, sends it to the server, then receives a response.
    \item \textit{Storage \& Processing:}
        The server stores the data item in the database.
\end{enumerate}
\begin{enumerate}[start=1,label={\Circled{T\arabic*}}]
    \item \textit{Handover:}
        The data source provides the data item and additional data (for membership testing) to the client.
    \item \textit{Monitoring:}
        The monitor downloads the database content to check for any misissued data items.
    \item \textit{Membership Testing:}
        The client runs an interactive verification algorithm with the server. The algorithm that outputs true or false depending on whether the data item of interest is stored in the database.
    \item \textit{Consistency Checking:}
        The client runs an interactive verification algorithm with the monitor and the server. The algorithm outputs true or false depending on whether there are any data items that have been removed from the database and that the is only one copy of the database.
\end{enumerate}

\utaggedpara{Example: Drug supply-chain auditing}
\Circled{U1} Pharmaceutical supply chain partners (data sources) submit their data to a public log (server/database), and
\Circled{U2} The data gets stored in a public log.
\Circled{T1} The consumer (client) purchases a medicine that contains the drug label (private data item), which also allows them to obtain the source-assisted information, e.g., in a form of a QR code,
\Circled{T2} The regulatory inspector (monitor) downloads the supply chain information to verify that there are no issues with the medicine,
\Circled{T3} The consumer checks that the supply chain information for the drug label is indeed stored in the public log, and
\Circled{T4} The consumer verifies that no supply chain information is missing and that there is only one copy of the public log with the help of the regulatory inspector.

\subsection{Requirements}

SPMT protocol has the following security requirements:
    \textit{Client privacy}, i.e., servers, monitors, and database hosts must not be able to gain any advantage in correctly identifying what the client's interest is; and
    \textit{Correctness}, i.e., if the system outputs true, then the data item of interest is indeed stored in the publicly monitored database, or any of the attacker's wrongdoings will be caught.

It should also fulfill the following non-security requirements:
    \textit{Low communication cost}, i.e., messages exchanged during data submission/membership testing should be constant with respect to the database size;
    \textit{Low computation cost}, i.e., client-/server-side computation during data submission/membership testing should be constant with respect to the database size;
    \textit{No client-side storage}, i.e., the protocol must not rely on client-side storage; and
    \textit{High scalability}, i.e., the proposed system should scale to large databases and handle large numbers of data submissions/client requests.

\utaggedpara{Example: Certificate Transparency}
When submitting a certificate to a CT log, a Certification Authority (CA) should not be kept waiting for minutes (low communication, computation cost).
Since millions of certificates are submitted each day~\cite{cloudflare_merkle_nodate}, the system should support a large number of CAs (scalability).
The client must not be required to submit their certificate during membership testing (client privacy), but must be able to efficiently (low communication, computation cost) know whether the CT log has stored the certificate (correctness) and should not be required to store any information (no client-side storage).
Moreover, since clients from all around the globe conduct membership testing, the system should be able to serve this large number of clients (scalability) without disrupting their browsing activity (low communication, computation cost).

\taggedpara{Non-requirements}
Preventing Denial-of-Service (DoS) attacks is not a requirement for an SPMT protocol, as malicious servers can always trivially launch such attack (e.g., by not replying to any client requests).

%% file: figures/spmt.tex
\begin{tikzpicture}[
    block/.style={
        draw, 
        rectangle, 
        rounded corners=10pt, 
        minimum width=4cm, 
        minimum height=2cm, 
        align=center,
        line width=1pt
    },
    database/.style={
        draw,
        cylinder,
        shape border rotate=90,
        minimum width=4cm,
        minimum height=2cm,
        aspect=0.25,
        align=center,
        line width=1pt
    },
    arrow/.style={
        -{Stealth[scale=1.2]},
        line width=1pt
    },
    dashedarrow/.style={
        -{Stealth[scale=1.2]},
        dashed,
        line width=1pt
    }
]

    \node[block, minimum width=10cm] (server) at (-2,3) {\large Server};
    
    \node[database] (db) at (9,3) {\large Database};
    
    \node[block] (source) at (-5,0) {\large Data\\\large Source};
    
    \node[block] (client) at (2,0) {\large Client};
    
    \node[block] (monitor) at (9,0) {\large Monitor};

    \draw[arrow] ([xshift=-0.1cm]source.north) -- ([xshift=-3.1cm]server.south) 
        node[midway, left, align=center] {\large \Circled{U1} Data\\\large Submission};
    \draw[arrow] ([xshift=-2.8cm]server.south) -- ([xshift=0.2cm]source.north);

    \draw[arrow] (server.east) -- (db.west) 
        node[midway, below, align=center, yshift=-0.0cm] {\large \Circled{U2} Storage\\ \large \& Processing};

    \draw[dashedarrow] (source.east) -- (client.west) 
        node[midway, above, solid] {\large \Circled{T1} Handover};

    \draw[dashedarrow] (db.south) -- (monitor.north) 
        node[midway, right, solid] {\large \Circled{T2} Monitoring};

    \draw[dashedarrow] ([xshift=-0.1cm]client.north) -- ([xshift=3.9cm]server.south)
        node[pos=0.5, left, align=center, solid] {\large \Circled{T3} Membership\\ \large Testing};
    \draw[dashedarrow] ([xshift=4.2cm]server.south) -- ([xshift=0.2cm]client.north);

    \draw[dashedarrow] (monitor.west) -- (client.east) 
        node[midway, above, align=center, solid] {\large \Circled{T4} Consistency\\ \large Checking};

\end{tikzpicture}

%% file: contents/04a-overview.tex
\section{Solution Intution} 
\label{sec:designoverview}

Supporting clients with source-assisted information does not make the SPMT protocol trivially overcome the issues of PMT.
A naive approach to conducting an SPMT protocol is to have the data source send the database to the client so that it can perform membership testing locally.
    Although this approach does not leak any data to the server, requiring every client to download the database seems unrealistic, especially if the database is extremely large.
    Moreover, the client's database must be updated every time it performs SPMT, which can be expensive, especially for clients with limited bandwidth (e.g., mobile phones).
Therefore, we do not consider this approach in the proposed system.

A technique proposed by prior work~\cite{meiklejohn_sok_2022} is to have the server immediately store the data item $\dds$ submitted by the data source and generate a proof that $\dds$ has been included in the database (aka \emph{proof of inclusion}), which is sent along with $\dds$ to the client via the data source.
    For example, by leveraging a Merkle Hash Tree (MHT), the server can produce a proof of inclusion that consists of a digital signature of the MHT root and the MHT inclusion path from the leaf corresponding to $\dds$ up to the root.
Since proof of inclusions can be verified locally, clients do not need to communicate with the server, enabling local and private membership testing.

However, generating a proof of inclusion requires the server and the database to process and store $\dds$ in near real-time.
    Therefore, this approach remains a theoretical discussion and has not been deployed in the real world, as it does not scale to modern, large-scale systems, where processing $\dds$ can take an unpredictable amount of time due to various reasons, e.g., the sheer size of data that require processing, complex pre-processing, and global distribution.

Moreover, this approach has a privacy drawback~\cite{meiklejohn_sok_2022}.  
    Since the proof of inclusion received by the client would most likely include an old MHT root, the client must check whether the database has not excluded any data up to the point of membership testing.
    To do so, the client transmits the old MHT root to the server and requests the server's root along with the sibling paths that differ between the two roots.
    With this information, the client reconstructs the root of the server and confirms that the database is not removing any data.  
    However, if a malicious server creates an MHT root that is updated only with $\dds$, sending this MHT root during this process will reveal the client's interest.

\subsection{Testing Passage of Time \& its issues}

The main objective of an SPMT protocol is to convince a client that a certain data item is stored in a database without violating the client's privacy.
    If we can design a system that is non-interactive, i.e., allows the client to conduct membership testing locally by utilizing the source-assisted information, then this system is by definition privacy-preserving, since there exists no information that can possibly be passed from the client to the server.
    Moreover, since clients no longer need to interact with the server, there is an added benefit for servers as they can focus their resources on accepting data items from data sources and storing them in the database.

One method that has been overlooked by previous work is to leverage \emph{the passage of time}.
 That is, if the server can ensure that it will store data items before a certain deadline, then the client can locally check whether enough time has passed to be convinced that the data item is stored in the database.
    To realize this, the server must be able to guarantee that it will (1) follow the pre-defined protocol (i.e., guarantee \emph{execution integrity}), and (2) store the data item in a database before a certain deadline (i.e., guarantee \emph{execution timeliness}).

Now, recall that in membership testing protocols, the server is considered to be actively malicious, where it may attempt to violate the guarantees shown above.
    Hardware-based trusted computing, e.g., Trusted Execution Environments (TEEs), have shown that upholding execution integrity under malicious environments is possible.
However, it is impossible to build a system that guarantees timeliness in malicious environments, even if the system leverages the execution integrity property of currently available TEEs~\cite{alder2023about}.
    This is because a malicious server can always delay packets that are sent to and from the TEE or temporarily halt TEE execution.

In addition, it is impossible to prevent or detect certain attacks without the client interacting with external parties.
    For instance, as some TEEs have restrictions in the amount of secure memory (e.g., Intel SGX), the database would be maintained \emph{outside} of the TEE where the adversary has full control over it.
        Therefore, since malicious database operators can alter the database content, the client must perform checks to verify that the server is not removing any data items from the database.
        Additionally, it is impossible for clients to detect split-view attacks without interacting with external parties.

\subsection{Our Approach: Testing for Sufficient Progress}

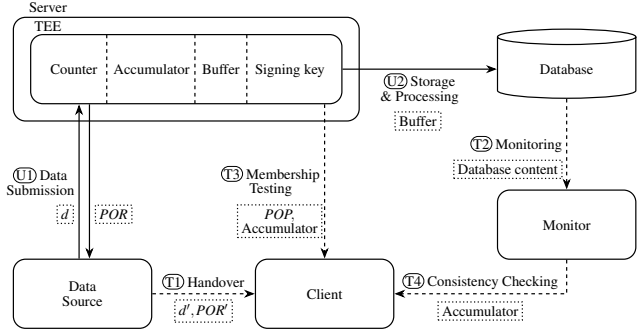
\begin{figure}[t]
    \centering
    \resizebox{\columnwidth}{!}{\input{figures/system_model}}
    \caption{\changed{Overview of our solution. Notations are listed in Table~\ref{table:notations}. Solid arrows represent Upload protocols and dashed arrows represent Testing protocols. Dotted boxes represent information exchanged during the interaction.}}
    \label{fig:system-model}
\end{figure}

    Since it is impossible to build a system that allows clients to remain strictly non-interactive, we take the viable subsequent approach where we introduce a number of carefully designed interactions.
    The first interaction is with the TEE during membership testing to check whether the TEE has not been interfered with by the server and is making \emph{sufficient progress} to store the data item in the database.
        There are different ways to realize this approach, and this work utilizes monotonically increasing counters stored within the TEE.
    The second interaction is with the monitor and the TEE to check whether the database is not excluding any data items and there are no malicious copies of the database.
        This is done by comparing the view of the TEE's and the monitor's view of the database content.

Figure~\ref{fig:system-model} provides an overview of our solution:
\Circled{U1}~Data sources submit data item $\dds$ to the TEE which is then stored in its buffer, and, in response, sends back a ``Proof Of Reception'' ($POR$) which asserts that the TEE has received $\dds$.
    $POR$ is a signature over $\dds$ and the counter value $\cntpor$.
\Circled{U2}~Once full, the TEE sends the buffer to the database.
    It simultaneously updates an accumulator (e.g., a hash chain) with the content of the buffer.
    Once updated, the TEE increments the counter.
\changed{\Circled{T1}~The client receives $POR'$ for the $\dds '$ it is interested in.}
\Circled{T2}~The monitor downloads the database content and updates its accumulator with the database content.
\Circled{T3}~During membership testing, the client requests a ``Proof Of Processing'' ($POP$), a signature over the current counter value, from the TEE to check whether the TEE has been processing the data items.
    $POP$ is a signature over an appropriate counter value $\cntpop$.
    If $POR'$ and $POP$ are correct and $\cntpor' < \cntpop$ is true, the client has assurance that ``the TEE has received and processed $\dds'$''.
\Circled{T4}~The client requests the accumulator from the TEE and the monitor.
    If the two accumulator values match (i.e., ``the TEE's and monitor's view of the database are consistent''), the client has assurance that $\dds'$ must be included in the database.

\taggedpara{Note}
In the actual design presented in Section~\ref{sec:designdetails}, $\cntpop$ is the counter associated with the hash-chain value of the monitor.
This is to accommodate cases where monitors may lag behind databases (see Section~\ref{sec:designdetails:freq} for details). 

%% file: figures/system_model.tex
\begin{tikzpicture}[
    % Define block styles
    block/.style={
        draw, 
        rectangle, 
        rounded corners=10pt, 
        minimum width=4cm, 
        minimum height=2cm, 
        align=center,
        line width=1pt
    },
    dashedblock/.style={
        draw, 
        rectangle, 
        dotted,
        align=center,
        line width=1pt
    },
    database/.style={
        draw,
        cylinder,
        shape border rotate=90,
        minimum width=4cm,
        minimum height=2cm,
        aspect=0.25,
        align=center,
        line width=1pt
    },
    arrow/.style={
        -{Stealth[scale=1.2]},
        line width=1pt
    },
    dashedarrow/.style={
        -{Stealth[scale=1.2]},
        dashed,
        line width=1pt
    },
    dashedline/.style={
        dashed,
        line width=1pt
    }
]

    % --- Nodes (Fixed Positioning) ---
    
    % Server: Centered at the top
    \node[block, minimum width=10cm, minimum height=3cm] (server) at (-2,8.5) {};
    \node at (-6,10.3) {\large Server};

    % TEE
    \node[block, minimum width=9cm] (tee) at (-2,8.5) {};
    \node at (-6,9.75) {\large TEE};

    \node at (-5.25,8.5) {\large Counter};
    \draw[dashedline] (-4.30,9.5) -- (-4.30,7.5);
    \node at (-3,8.5) {\large Accumulator};
    \draw[dashedline] (-1.75,9.5) -- (-1.75,7.5);
    \node at (-1,8.5) {\large Buffer};
    \draw[dashedline] (-0.25,9.5) -- (-0.25,7.5);
    \node at (1,8.45) {\large Signing key};
    
    % Database: Top Right
    \node[database] (db) at (9,8.5) {\large Database};
    
    % Data Source: Bottom Left
    \node[block] (source) at (-5,2) {\large Data\\\large Source};
    
    % Client: Bottom Center
    \node[block] (client) at (2,2) {\large Client};
    
    % Monitor: Bottom Right
    \node[block] (monitor) at (9,4) {\large Monitor};

    % --- Connections ---

    % 1. Data Submission (Server <-> Data Source)
    \draw[arrow] ([xshift=-0.1cm]source.north) -- ([xshift=-3.1cm]tee.south) 
        node[midway, left, align=center] {\large \Circled{U1} Data\\\large Submission};
        \node[dashedblock] at (-5.5,4.3) {\large $d$};
    \draw[arrow] ([xshift=-2.8cm]tee.south) -- ([xshift=0.2cm]source.north);
        \node[dashedblock] at (-4.1,4.3) {\large $POR$};

    % 2. Storage & Processing (tee -> Database)
    \draw[arrow] (tee.east) -- (db.west) 
        node[midway, below, align=center, yshift=-0.0cm] {\large \Circled{U2} Storage\\ \large \& Processing};
        \node[dashedblock] at (4.7,7) {\large Buffer};

    % 3. Handover (Data Source -> Client)
    \draw[dashedarrow] (source.east) -- (client.west) 
        node[midway, above, solid] {\large \Circled{T1} Handover};
        \node[dashedblock] at (-1.5,1.5) {\large $d', POR'$};

    % 4. Monitoring (Database -> Monitor)
    \draw[dashedarrow] (db.south) -- (monitor.north) 
        node[midway, left, solid] {\large \Circled{T2} Monitoring};
        \node[dashedblock] at (7.3,5.6) {\large Database content};

    % 5. Membership Testing (tee <-> Client)
    \draw[dashedarrow] ([xshift=4cm]tee.south) -- ([xshift=0.0cm]client.north)
        node[pos=0.5, left, align=center, solid] {\large \Circled{T3} Membership\\ \large Testing};
        \node[dashedblock] at (0.7,4.1) {\large $POP,$ \\ \large Accumulator};

    % 6. Consistency Checking (Monitor -> Client)
    \draw[dashedarrow] (monitor.south) -- (monitor.south |- client.east) -- (client.east) 
        node[midway, above, align=center, solid] {\large \Circled{T4} Consistency Checking};
        \node[dashedblock] at (6.5,1.5) {\large Accumulator};

\end{tikzpicture}

%% file: contents/04b-design.tex
\section{Solution Details} 
\label{sec:designdetails}

\begin{table}[t]
\centering
\footnotesize
\caption{Notations \& Definitions}
\label{table:notations}
\begin{tabular}{@{}ll@{}}
\toprule
Notation                   & Definition                          \\ \midrule
$pk_{DS},~pk_{T},~pk_{M}$ & Public key of Data Source, TEE, or Monitor \\
$sk_{DS},~sk_{T},~sk_{M}$ & Private key of Data Source, TEE, or Monitor \\
$\dds$                       & Data item submitted by the data source  \\
$\dds '$                       & \changed{Data item the client is interested in}  \\
$sig_{\dds}$                       & Digital signature over $\dds$ with $\skds$  \\
$cnt$                      & Counter value maintained by TEE \\
$\cntpor$                    & Counter value at the time of data submission \\
$\cntpop$                    & Counter value at the time of membership testing \\
\texttt{buffer}                    & Buffer to store $\dds$ submitted by data sources \\
$n$                      & Maximum number of $\dds$ stored in \texttt{buffer} \\
$ACK$                    & Acknowledgment sent from database \\
$POR$                    & Proof Of Reception \\
$POR'$                    & \changed{$POR$ of $\dds '$ which the client is interested in} \\
$POP$                    & Proof Of Processing \\
$HC_T$                    & TEE's Hash Chain value \\
$List_{HC_T}$                    & List of past $(HC_T, cnt)$ tuples \\
$HC_M$                    & Monitor's Hash Chain value \\
$sig_{HC_M}$                       & Digital signature over $HC_M$ with $\skm$  \\
$report$                    & Attestation report \\
$APK$                    & A well known Attestation Public Key \\
$EM$                    & Expected TEE measurement value \\
\bottomrule
\end{tabular}
\end{table}

\input{figures/sequence_diagram}

\changed{\sys consists of the following protocols: \Circled{0} Initialization, \Circled{U1} Data submission, \Circled{U2} Storage \& Processing, \Circled{T1/2} Handover and Monitoring, and \Circled{T3/4} Membership testing and Consistency checking.}
The sequence diagrams and algorithms of the data submission / storage \& processing protocols and the membership testing \& consistency checking protocol are depicted in Figures~\ref{fig:submission_processing} and \ref{fig:membership_testing}, respectively.

\taggedpara{\Circled{0} Initialization}
The data source generates a set of key pairs $\pkds$ and $\skds$.
The server initializes the TEE that generates the key pair $\pks$ and $\sks$, sets $cnt = 0$, \texttt{buffer} and $List_{HC_T}$ to $\emptyset$, and contacts the attestation infrastructure to obtain $report$. 
$report$ is a signature over the platform / code integrity measurements and $\pks$, allowing attesting parties to have the assurance that digital signatures that can be verified with $\pks$ must be generated via the TEE that runs a certain code.
The client obtains $APK$ and $EM$ from a trustworthy source, e.g., by contacting attestation infrastructures and source code repositories.

\taggedpara{\Circled{U1} Data submission}
The data source generates data item $\dds$ and a digital signature $sig_{\dds} = Sign_{\skds}(\dds)$, and sends them to the TEE.
After receiving the two pieces of data, the TEE first verifies $sig_{\dds}$ over $\dds$.
We assume that the TEE is given a set of trusted data source certificates which is used for verification.
The TEE subsequently appends $\dds$ to its \texttt{buffer}.
It then sends the data source a $POR = \{Sign_{\sks}(\dds || \cntpor || report), \cntpor, report\}$, where $\cntpor$ is the counter value at the point of receiving $\dds$, $report$ the attestation report, $||$ a concatenation operation, and $\sks$ the TEE's private key.

\taggedpara{\Circled{U2} Storage and processing} \label{sec:designdetails:data_processing}
This protocol is triggered when \texttt{buffer} is full.
The TEE first exports \texttt{buffer} to the database for storage.
While the database processes and stores the buffer, the TEE updates $HC_T$ with the content of the buffer by continuously calculating $HC_T = HC(HC_T || \dds)$ for each $\dds$ in \texttt{buffer}.
The database sends the TEE an $ACK$ indicating that it has finished storing the buffer, which then triggers the TEE to add the $(HC_T, cnt)$ tuple to $List_{HC_T}$ and increment $cnt$.

Note that the data processing protocol is a \emph{blocking} execution, i.e., the TEE cannot receive any new $\dds$ during this period.
    This is necessary to maintain the total ordering of $\dds$.
    Although this protocol executes relatively fast, we optimize this design in our implementation for even faster processing.

\taggedpara{\Circled{T1/2} Handover and monitoring}
\changed{Later on, the client receives $\dds'$, which is one of the data items issued by the data source along with its $POR$. Note here that both pieces of information does not necessarily need to be sent directly from the data source and can be distributed via third party distributors, e.g., content distribution networks.}
The monitor downloads the database and updates $HC_M$ by calculating $HC_M = HC(HC_M || \dds)$ for each $\dds$ in the database.

\taggedpara{\Circled{T3/4} Membership testing and consistency check}
The client extracts $\cntpor$ and $report$ from $POR'$ and checks:

\vspace{1mm}
\begin{itemize}
    \item Whether $report$ can be verified using $APK$.
    \item Whether the code integrity measurement included in $report$ matches the expected value $EM$.
    \item Whether $POR'$ can be verified using $\pks$ which is extracted from $report$. 
    \item Whether $\dds'$ included in $POR'$ matches the data item it received from the data source.
\end{itemize}

If the above check fails at any point, the client discards $\dds'$ and deems the data source malicious.
The client then requests the monitor's hash chain value $HC_{M}$ and a digital signature over the value $sig_{HC_M}$ and forwards it to the TEE.
After receiving this value, the TEE first verifies whether the signature can be verified using the monitor's public key.
The TEE then retrieves the counter value associated with $HC_{M}$ stored in $List_{HC_T}$, which is denoted as $\cntpop$.
The TEE then generates $POP = \{Sign_{\sks}(\cntpop), \cntpop\}$, where $\cntpop$ is the counter value mentioned above and $\sks$ the private key of the TEE, and sends it to the client.
Subsequently, the client checks whether $POP$ can be verified using $\pks$.
If this check fails or the TEE sends an error, the client discards $POP$ and deems the server malicious.

The client then verifies whether $\cntpor' < \cntpop$.
If the above check passes, then the client has assurance that the TEE has processed the buffer that includes $\dds'$ \emph{and} that there is a point in time where the TEE's and monitor's view of the database that includes $\dds'$ is consistent, i.e., there were no data items removed from the database at that time and the monitor has $\dds'$ in its possession.
If the check fails, this means that the TEE has not yet processed $\dds'$ or the monitor has not pulled the latest content from the database.
    In this case, the client is recommended to conduct the membership test at a later time or contact other monitors in the hope that they are more synchronized with the database.

\subsection{Design Considerations}
    In our design, we must consider the following aspects.

\taggedpara{Database acknowledgments}
Once the database finishes storing the content of \texttt{buffer}, it sends an $ACK$ to the TEE.
    However, since the database is hosted on a potentially malicious server, the adversary can forge an $ACK$.
    For instance, a malicious server can send multiple $ACK$s in an attempt to increment the $cnt$ even though no data was being processed or stored.
    Therefore, the TEE must discard any $ACK$s that were sent outside of the storage \& processing protocol and only increment $cnt$ once, even if it receives multiple $ACK$s during a single storage \& processing phase.

One might ask why we include $ACK$ in the data processing protocol if it can be forged.
    The reason is that, under legitimate circumstances, $ACK$ allows the TEE to smooth out the rate at which it receives $\dds$.
    Assume that the TEE receives a burst of data submissions and that the speed at which the database stores data items is relatively stable.
    This allows $ACK$s to be sent out at a similarly stable rate.
    As a result, TEEs can even out the burst of incoming submissions, as responses for $\dds$ received during data processing would be delayed.

\taggedpara{Different data reception frequencies} \label{sec:designdetails:freq}
In an ideal scenario, the rate at which the TEE and the monitor receive data is the same, i.e., once the TEE submits \texttt{buffer} to the database, the monitor is immediately updated with the \texttt{buffer} content.
    However, this requires the monitor to constantly poll for new updates, which puts a substantial load on the database.
    Moreover, since there may be multiple monitors, it is not realistic to require all monitors to be in sync with the TEE.
    Therefore, it is more realistic to assume that the rate at which monitors are updated with data items is slower compared to the TEE's.
    This is the reason why we require the TEE to keep a list of previous $HC_T$ as $List_{HC_T}$, in the hope that one of them matches the one provided by the monitor that the client is contacting.
    Additionally, this is why the actual design deviates slightly from the one described in Section~\ref{sec:designoverview}.
    However, this raises the question: how many $HC_T$ from the past is the TEE required to store?

\taggedpara{Determining optimal list size}
It is not ideal to store the entire list of $(HC_T, cnt)$ tuples generated in the past, as secure storage may be limited in some TEEs (e.g., Intel SGX).
    Therefore, it is important to know the theoretical upper limit of the number of tuples that are strictly necessary to be stored, which we denote as $N$.
    Assume that the TEE receives data items at a frequency of $freq_T$ per second and that the monitor pulls the latest database content at a minimum frequency of $freq_M$ per second (since there may be multiple monitors, we must match the frequency to the slowest monitor).
    Assuming that the frequencies are stable, $N$ can be calculated as $N = \lceil freq_{T} / freq_{M} \rceil + 1$. 
    The intuition behind this equation is that the TEE must store the monitor's last-known hash chain value \emph{in addition} to the at most $\lceil freq_{T} / freq_{M} \rceil$ newer values.

%% file: figures/sequence_diagram.tex
\begin{figure}[th!]
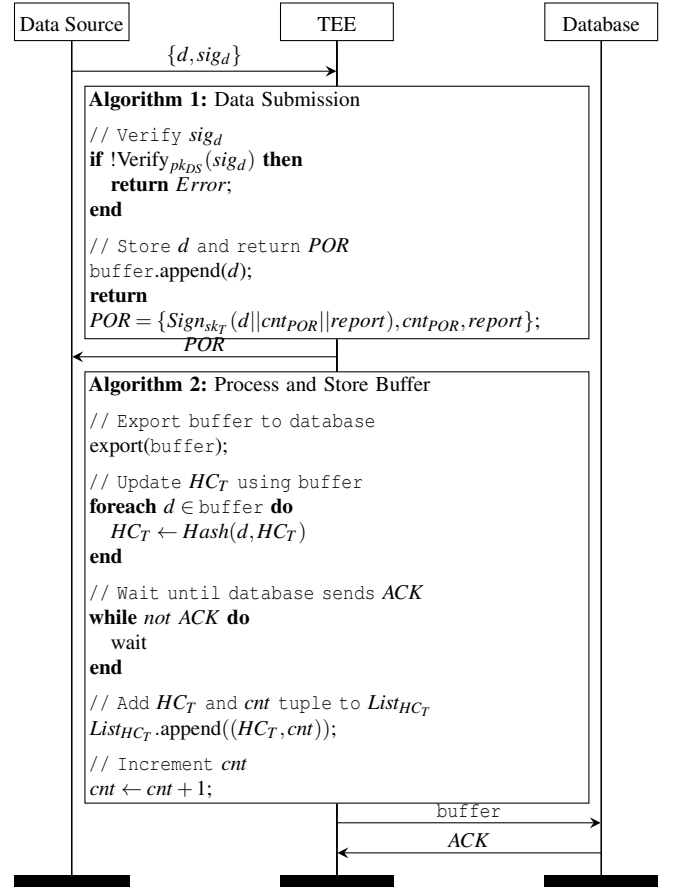

    \centering
    \begin{msc}[small values, draw frame=none, msc keyword=, head top distance=5mm, foot distance=0mm, instance distance=20mm, environment distance=10mm, level height=2mm, arrow scale=2.0, instance width=15mm, left inline overlap=15mm, right inline overlap=15mm, /tikz/font=\footnotesize, /tikz/line width=0.4pt, /msc/line width=0.4pt, label distance=0.4ex, action height=4mm]{ }
    	\declinst{di}{}{Data Source}
		\declinst{e}{}{TEE}
		\declinst{db}{}{Database}
  
		\mess{$\{\dds,sig_{\dds}\}$}{di}{e}
		\nextlevel[1]
		\action[action width=67mm]{\raggedright
            \textbf{Algorithm 1:} Data Submission\\
            \vspace{0.5em}
            \texttt{// Verify $sig_{\dds}$}\\
            \textbf{if} !Verify$_{\pkds}(sig_{\dds})$ \textbf{then} \\
            \quad \textbf{return} $Error$;\\
            \textbf{end}\\
            \vspace{0.5em}
            \texttt{// Store $\dds$ and return $POR$}\\
            \texttt{buffer}.append($\dds$); \\
            \textbf{return} $POR = \{Sign_{\sks}(\dds || \cntpor || report), \cntpor, report\}$;\\
        }{e}
		\nextlevel[18]
		\mess{$POR$}{e}{di}
		\nextlevel[1]
		\action[action width=67mm]{\raggedright
            \textbf{Algorithm 2:} Process and Store Buffer\\
            \vspace{0.5em}
            \texttt{// Export \texttt{buffer} to database}\\
            export(\texttt{buffer}); \\
            \vspace{0.5em}
            \texttt{// Update $HC_T$ using \texttt{buffer}}\\
            \textbf{foreach} $d \in \texttt{buffer}$ \textbf{do} \\
            \quad $HC_T \gets Hash(d, HC_T)$\\
            \textbf{end}\\
            \vspace{0.5em}
            \texttt{// Wait until database sends $ACK$}\\
            \textbf{while} $not~ACK$ \textbf{do} \\
            \quad wait\\
            \textbf{end}\\
            \vspace{0.5em}
            \texttt{// Add $HC_T$ and $cnt$ tuple to $List_{HC_T}$}\\
            $List_{HC_T}.\text{append}((HC_T, cnt))$;\\
            \vspace{0.5em}
            \texttt{// Increment $cnt$}\\
            $cnt \gets cnt + 1$;\\
        }{e}
		\nextlevel[30]
		\mess{\texttt{buffer}}{e}{db}
		\nextlevel[2]
        \mess{$ACK$}{db}{e}
    \end{msc}
    \caption{
    \changed{Upload (Data submission and Storage \& Processing) protocols. These protocol run for every $\dds$ submitted by a data source.}}
    \label{fig:submission_processing}
\end{figure}

\begin{figure}[th!]
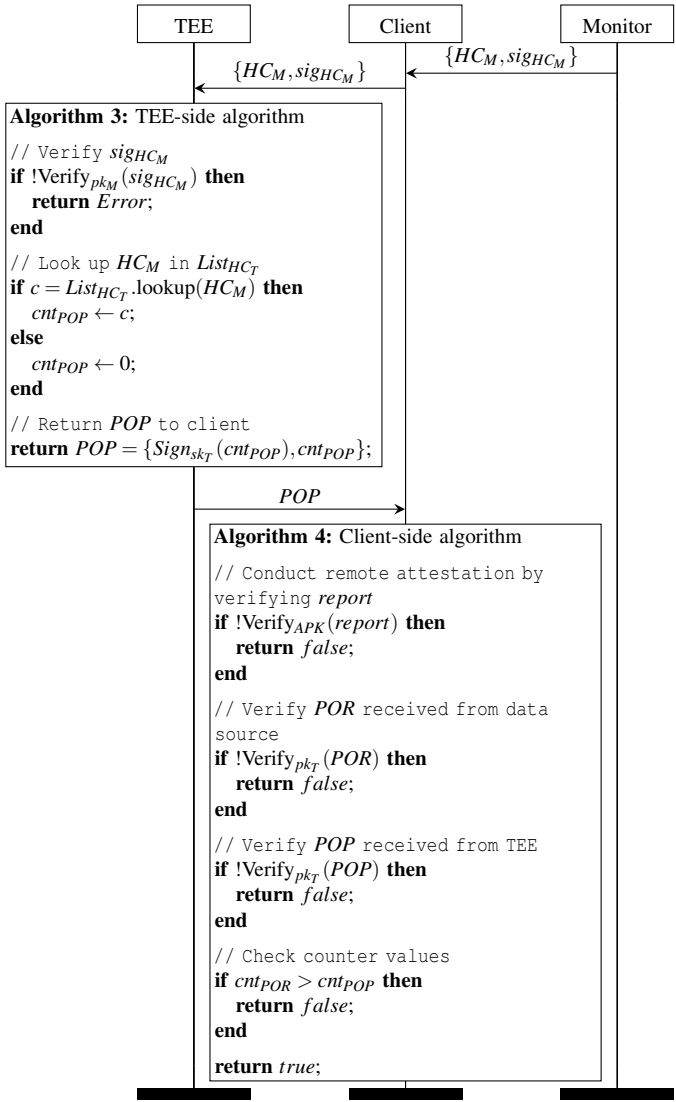

    \centering
    \begin{msc}[small values, draw frame=none, msc keyword=, head top distance=5mm, foot distance=0mm, instance distance=13mm, environment distance=10mm, level height=2mm, arrow scale=2.0, instance width=15mm, left inline overlap=15mm, right inline overlap=15mm, /tikz/font=\footnotesize, /tikz/line width=0.4pt, /msc/line width=0.4pt, label distance=0.4ex, action height=4mm]{ }
		\declinst{e}{}{TEE}
		\declinst{c}{}{Client}
		\declinst{m}{}{Monitor}
        
		\mess{$\{HC_{M}, sig_{HC_{M}}\}$}{m}{c}
		\nextlevel[1]
		\mess{$\{HC_{M}, sig_{HC_{M}}\}$}{c}{e}
		\nextlevel[1]
		\action[action width=50mm]{\raggedright
            \textbf{Algorithm 3:} TEE-side algorithm\\
            \vspace{0.5em}
            \texttt{// Verify $sig_{HC_M}$}\\
            \textbf{if} !Verify$_{\pkm}(sig_{HC_M})$ \textbf{then} \\
            \quad \textbf{return} $Error$;\\
            \textbf{end}\\
            \vspace{0.5em}
            \texttt{// Look up $HC_M$ in $List_{HC_T}$}\\
            \textbf{if} $c = List_{HC_T}.\text{lookup}(HC_M)$ \textbf{then} \\
            \quad $\cntpop \gets c$;\\
            \textbf{else} \\
            \quad $\cntpop \gets 0$;\\
            \textbf{end}\\
            \vspace{0.5em}
            \texttt{// Return $POP$ to client}\\
            \textbf{return} $POP = \{Sign_{\sks}(\cntpop), \cntpop\}$;\\
        }{e}
		\nextlevel[27]
		\mess{$POP$}{e}{c}
		\nextlevel[1]
		\action[action width=52mm]{\raggedright
            \textbf{Algorithm 4:} Client-side algorithm\\
            \vspace{0.5em}
            \texttt{// Conduct remote attestation by verifying $report$}\\
            \textbf{if} !Verify$_{APK}(report)$ \textbf{then} \\
            \quad \textbf{return} $false$;\\
            \textbf{end}\\
            \vspace{0.5em}
            \texttt{// Verify $POR$ received from data source}\\
            \textbf{if} !Verify$_{\pks}(POR)$ \textbf{then} \\
            \quad \textbf{return} $false$;\\
            \textbf{end}\\
            \vspace{0.5em}
            \texttt{// Verify $POP$ received from TEE}\\
            \textbf{if} !Verify$_{\pks}(POP)$ \textbf{then} \\
            \quad \textbf{return} $false$;\\
            \textbf{end}\\
            \vspace{0.5em}
            \texttt{// Check counter values}\\
            \textbf{if} $\cntpor > \cntpop$ \textbf{then} \\
            \quad \textbf{return} $false$;\\
            \textbf{end}\\
            \vspace{0.5em}
            \textbf{return} $true$;\\
        }{c}
		\nextlevel[36]
    \end{msc}
    \caption{\changed{Membership testing \& Consistency checking protocol. This protocol run only for a limited selection of data items which the client requested from the data source.}}
    \label{fig:membership_testing}
\end{figure}

%% file: contents/06-evaluation.tex
\section{Performance Evaluation} 
\label{sec:eval}

\noindent\textbf{Implementation. }
Our implementation focuses on testing the applicability of our proposed system in practice, as well as evaluating real-life performance. We aim to examine the latency and throughput behaviors for different numbers of concurrent clients. To accommodate the high concurrency required to handle large numbers of simultaneous clients, we chose to implement our TEE functionality in Go (1.5K LoC). \sys is based on several configurable parameters, including the buffer size for incoming submissions and the size of the hash chain history maintained by the TEE. The size of the hash chain history has no practical impact on latency and throughput measurements, as it is implemented as a hash table, resulting in constant-time lookups.

We leverage Go's standard library to compute signatures between different entities using ECDSA over the curve SECP256R1, with SHA256 as the hashing primitive, although our solution is not tied to a specific set of cryptographic primitives. SQLite3 is used as an in-memory database to store incoming items on the database host.

We leverage AMD-SEV as a TEE, although our protocol does not depend on any SEV-specific features and could also be deployed on Intel SGX, Intel TDX, or upcoming TEEs such as Arm CCA, \changed{and anticipate that similar performance numbers would be demonstrated}.
We utilize the Azure ecosystem to simplify the AMD-SEV attestation process by leveraging Microsoft's Azure Attestation Service.

We benchmark the proposed solution on an Azure Cloud environment. The TEE functionalities run on a 16-core AMD-SEV SNP machine (Azure VM type DC16ads v5), while a conventional 32-core VM (D32ads v6) is used to simulate data sources performing data submissions and clients performing membership testing. The database is implemented on an 8-core D8ads v6 VM. All machines run Ubuntu 24.04, with the client and database machines using kernel version 6.11, and the TEE running version 6.8.

\utaggedpara{Simplifying assumptions}
For our implementation and subsequent evaluation, we assume that all public keys have been distributed to the relevant parties beforehand. As such, data sources and clients are considered to have obtained the public key of the TEE, and the TEE is assumed to have obtained the public keys of any potential data sources and monitors involved in the protocol. In practice, these keys would be either distributed by PKI or extracted from the TEE's attestation report.
The existing CT monitoring infrastructure already shows that monitoring large-scale public logs is practical, thus we believe that the same can be applied to our system and therefore did not implement the monitor.
Although network latency between the client and monitor would slightly increase the membership testing latency, this increase is constant and, since it is deployment-dependent, it was excluded from measurements.

\utaggedpara{Implementation optimizations}
The proposed protocol can be implemented in multiple ways. As discussed in Section~\ref{sec:designdetails:data_processing}, one possible design involves blocking clients while the server processes a given batch of submissions. To optimize the throughput and latency of our implementation, we avoid blocking clients and immediately acknowledge submissions. The returned counter during data submission indicates the batch index to which an item was added. The server keeps track of how many batches have been processed so far and returns that counter to clients as part of the $POP$. This allows the latency and throughput of our implementation to be independent of the server's batch size.

\subsection{Latency Evaluation} \label{sec:eval:latency}

We report the average network round-trip time (RTT) between the different entities in our deployment, measured in milliseconds, and averaged over 100 measurements using ICMP echo requests. The TEE and the client machine (used to simulate data sources and membership testing clients) have an average network RTT of $0.5\pm1.2$ms, while the TEE and the database host have an RTT of $0.9\pm0.5$ms. As both Upload and membership testing \& consistency checking protocols are independent of database size, the following measurements apply to small and large databases.

\begin{figure}[t!]
    \centering
    \includegraphics[width=0.4\textwidth]{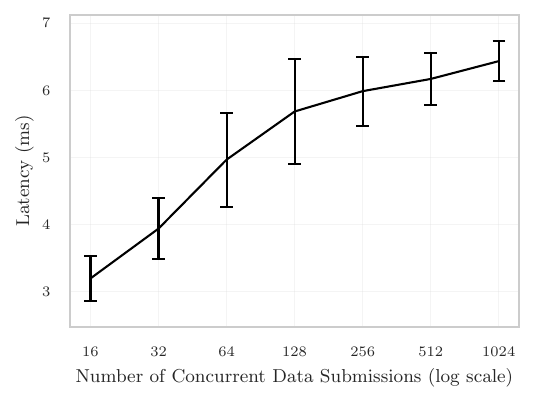}
    \caption{Upload latency evaluation for various concurrent data sources. We achieve a data submission latency of around 6.5ms when measured over 1024 concurrent clients.}
    \label{fig:latency-submission}
\end{figure}

\taggedpara{Upload latency} \label{subsec:datasublatency}
We measure the Upload (data submission and storage \& processing) latencies for various numbers of concurrent data sources. Data sources submit a single 4.9 KB-size data item at a time, roughly corresponding to the size of average TLS certificates in PEM format. Note that our implementation supports any data size.
Our measurements are averaged over 50 repeated iterations for varying numbers of concurrent data submissions. We benchmark data submission latencies for $\{16,32,64,128,256,512,1024\}$ concurrent data submissions. We filter outliers using the 1.5 Interquartile Range (IQR) rule. Measurements are visualized in Fig.~\ref{fig:latency-submission}. Of a total of $350$ data points, we filtered out 6 outliers (1.7\%). Our results show that \sys can handle a large number of concurrent data submissions with a latency of around 6.5 ms.

\begin{figure}[t!]
    \centering
    \includegraphics[width=0.4\textwidth]{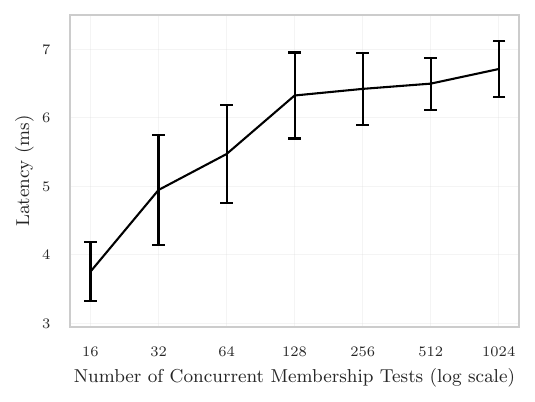}
    \caption{Membership testing and Consistency checking latency evaluation for various concurrent clients. Latencies include cryptographic verifications of the $POR$, $POP$, and TEE attestation report. We achieve a latency of under 7ms.}
    \label{fig:latency-mt}
\end{figure}

\taggedpara{Membership testing and Consistency checking latency}
The membership testing and consistency checking process includes several steps, as described in Fig.~\ref{fig:membership_testing}. We aim to quantify the latency of the entire process, which consists of verifying $POR$, extracting and validating $report$, fetching $POP$, and verifying $POP$\footnote{$POR$ can be verified before conducting the membership testing process. We include it in the latency measurement to provide the worst-case latency.}. We assume that at the start of the process the client has already received $POR$ from the data source and $\{HC_M, sig_{HC_M}\}$ from the monitor. Measurements are averaged over 50 repeated iterations for each concurrent client count ($\{16,32,64,128,256,512,1024\}$). We visualize our results in Fig.~\ref{fig:latency-mt}. Of the $350$ data points, we filtered out $12$ outliers ($3.4\%$). Results show \sys can handle large number of concurrent membership testing with minimal latency of around 7 ms.

\taggedpara{Latency breakdown}
We measure each operation on the server side, using the same setup that was used for the latency evaluation. We use $512$ concurrent data sources that submit exactly one certificate and measure the latencies of various server-side components associated with data submissions, averaged over 50 trials.

\utaggedpara{Data submission}
The function responsible for handling data submission requests takes $0.29 \pm 0.22$ ms and is primarily dominated by buffer operations, which require holding a lock ($0.13 \pm 0.20$) and cryptographic operations such as signing and verifying submission signatures ($0.11 \pm 0.04$).

\utaggedpara{Processing \& storage}
The server asynchronously processes data every 32 submissions, forwarding them to the database. Although this does not impact client submission latencies, we include a breakdown of the processing phase for completeness. The processing phase takes, on average, $2.66 \pm 1.16$ ms and is dominated by the cost of sending items to the database over the network ($1.03 \pm 0.53$). Waiting for the database to commit and acknowledge the sent items takes $0.92 \pm 0.18$ ms. 

\utaggedpara{Membership testing \& Consistency checking}
On the server side, this process is dominated primarily by the verification of $sig_{HC_M}$ and the signing operation while generating $POP$. These operations take $0.11 \pm 0.03$ and $0.07\pm 0.07$ ms, respectively.
On the client side, TCP network socket operations ($5.11 \pm 0.41$ ms) have the greatest impact on latency, followed by attestation verification procedures ($0.25 \pm 0.00$ ms).

\subsection{Throughput Evaluation} \label{sec:eval:bandwidth}

We evaluate the maximum throughput of \sys by measuring the maximum number of requests processed when considering $C$ clients submitting back-to-back requests over 5 seconds while varying $C$. The results are then normalized by dividing the resulting submission rate by 5 and the number of logical cores of the server machine, resulting in submissions per second per core. Outliers are filtered with the same methodology as used for latency benchmarks (1.5 IQR).

\begin{figure}[t!]
    \centering
    \includegraphics[width=0.4\textwidth]{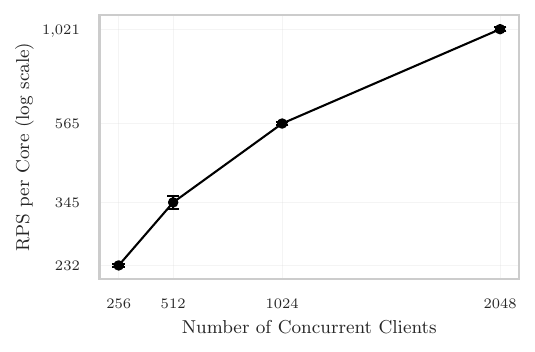}
    \caption{Maximal upload throughput, measured per second per core. On a 16-core machine, our implementation can handle approximately 16336 data submissions/sec from 2048 concurrent clients. The submission rate is measured by averaging the number of submissions made by concurrent data sources that continuously submit data over a 5-second period.}
    \label{fig:eval-throughput-submission}
\end{figure}

\taggedpara{Upload throughput}
We report a throughput of up to 1021 requests/sec/core for 2048 concurrent clients (see Fig.~\ref{fig:eval-throughput-submission}).

\begin{figure}[t!]
    \centering
    \includegraphics[width=0.4\textwidth]{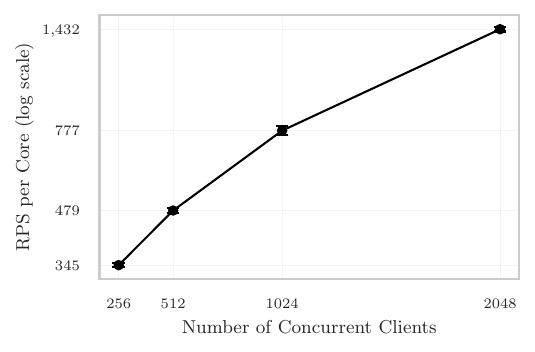}
    \caption{Maximal membership testing throughput for $POP$ requests, measured per second per core. Considering 2048 concurrent clients, using a 16-core server, our implementation can handle approximately 22912 $POP$ requests/sec. The submission rate is measured by averaging the number of submissions made by concurrent clients that continuously submit $POP$ requests within a 5-second interval.}
    \label{fig:eval-throughput-audit}
\end{figure}

\taggedpara{Membership testing throughput}
We achieve a throughput of approximately 1432 requests/sec/core for 2048 concurrent clients (see Fig.~\ref{fig:eval-throughput-audit}).

\subsection{Computation Evaluation}
\taggedpara{Server-side computation}
The only computation necessary for the server is to verify $sig_{\dds}$ and generate $POR$ for the data submission protocol and verify $sig_{HC_M}$, retrieve $\cntpop$, and generate $POP$.
    As these operations are independent of the number of entries in the database, the server-side computation is constant.

\taggedpara{Client-side computation}
The only computation necessary for the client is to verify $report, POR, POP$ and verify the relationship between $\cntpor$ and $\cntpop$.
    As these operations are independent of the number of entries in the database, the client-side computation is constant.

\subsection{Certificate Transparency Case Study}
All presented latency and throughput results were obtained using a database containing 3.3 million certificates. %

\taggedpara{Upload}
Approximately 460,000 certificates are submitted to CT logs per hour (2024, \cite{ostertag2024anomaly}), which is roughly 128 certificates/sec.
Since our achieved throughput capacity is 1021 requests/sec/core, this is clearly sufficient to meet real-world requirements. 

\taggedpara{Membership testing}
Crinchton et al. reported that a user visits on average 163 distinct web pages per day (2023,~\cite{crichton2023how}).
Making a conservative assumption where every web page's  certificate visited by the user is logged by a CT log, the user would need to perform 163 membership tests per day, which results in an average of approximately 0.002 web pages per second.
For 2048 clients, this will result in approximately 4.1 membership tests per second, and given that \sys achieves 1432 requests/sec/core, the system is well capable of handling the load.

\section{Security Analysis} \label{sec:eval:security}

\subsection{Client Privacy}
We provide a game-based definition for client privacy.

\taggedpara{Privacy game}
    Let $A$ be the adversary and $C$ be the challenger.
    $A$ corrupts the server, allowing it to observe any data and modify protocol execution arbitrarily.
    $A$ corrupts the monitor, allowing it to observe execution and messages.
    $A$ corrupts the TEE, allowing it to leak data used in any non-cryptographic processing. 

    $A$ is given access to the Data Submission oracle and it populates the database with freely-chosen data using the oracle.
    $A$ then chooses two data items $d_0$ and $d_1$, populates the database with both items, and sends them to $C$.
    $C$ picks a random bit $b \leftarrow \{0,1\}$ and executes the Membership Testing and Consistency Checking protocol for $d_b$.
    $A$ makes guess $b' = \{0,1\}$.
    $A$ wins the game if $b = b'$.

\taggedpara{Definition}
    An SPMT protocol is private if no $A$ exists that can win the game at a probability better than random guessing.

\taggedpara{Claim}
    \sys is private for an unbounded number of protocol runs.

\taggedpara{Proof sketch}
    Messages exchanged between the client and the server, and between the client and the monitor, are independent of $\dds$ and the database.
    Therefore, $A$ observes the same messages for $d_0$ and $d_1$, thus making the protocol executions indistinguishable.

\subsection{Correctness}
For correctness, we provide a systematic, but less formal analysis. Correctness means that (1) if the membership testing and consistency checking algorithm outputs true, then the data item must be stored in the database, and (2) the adversary gets caught of its wrongdoings, if any. We analyze the correctness guarantee of \sys by first defining an ideal scenario that clearly provides correctness and bringing it closer to our solution step by step \changed{to demonstrate that our solution provides correctness. Due to page restrictions, the full analysis is shown in Appendix~\ref{appendix:correctness}.}

%% file: contents/07-application.tex
\section{Applications} \label{sec:applications}

\subsection{\sys-supported use-cases}
\taggedpara{CT log auditing}
Certificate Transparency (CT) logs store certificates that are issued by Certification Authorities (CAs) and are used by domain owners to check whether there are any maliciously issued certificates under their domain.
In the context of CT, web browsers (\textit{clients}) are interested in checking whether website certificates are issued by valid CAs (\textit{data sources}) and are stored in valid CT logs (\textit{server/database}).
In this example, browsers would engage in an SPMT protocol with the CT log utilizing Signed Certificate Timestamps (\textit{source-assisted information}) in addition to the certificate, checking whether the CT log has indeed stored the certificate with the help of CT monitors (\textit{monitors}).

\taggedpara{Supply chain auditing}
As the risks associated with supply chain disruptions continue to escalate, it is essential for consumers to verify if the products they purchase provide transparency regarding their supply chain, enabling third-party inspection.
For instance, US pharmaceutical companies must adhere to the Drug Supply Chain Security Act (DSCSA), which mandates that data from supply chain partners (\textit{data sources}) be recorded in public logs (\textit{server/database}) and subsequently inspected by regulatory agencies (\textit{monitors}).
In this example, consumers (\textit{clients}) would engage in an SPMT protocol with the public log utilizing information from the supply-chain partners (\textit{source-assisted information}) provided during the purchase of the drug in addition to the drug label, checking whether the public log has indeed stored the supply-chain information with the help of regulatory agencies.
This enables consumers to verify if companies comply with regulations, without revealing the purchased product.

\taggedpara{Data provenance verification}
In light of the emergence of generative AI, the barrier to entry to creating ``deepfaked'' photos and videos has decreased significantly.
Consequently, it is important for individuals to have the ability to verify the authenticity of the photos and videos they encounter by checking if they are publicly logged, thereby enabling third-party inspection.
For instance, the Coalition for Content Provenance and Authenticity (C2PA) is an initiative that aims to address the prevailing misleading information in the public domain.
A journalist (\textit{data source}) captures an image of a politically sensitive incident and records it in a public log (\textit{server/database}) hosted by the Content Authenticity Initiative (CAI), which adheres to the C2PA specification and is then inspected by fact-checking organizations (\textit{monitors}).
    In this example, users (\textit{clients}) that view this photo over the Internet would interact with the CAI public log leveraging the SPMT protocol with the information associated with the photo (\textit{source-assisted information}) in addition to the photo itself, verifying whether the CAI public log has stored the photo with the help of fact-checking organizations.

\taggedpara{Public document auditing}
Inspecting publicly accessible documents released by governments, corporations, and other entities enables watchdog organizations to oversee their activities, providing a crucial layer of oversight for powerful entities.
For example, governments (\textit{data sources}) are required to publish budget documents on public online records (\textit{server/database}) so that they can be inspected by watchdog organizations (\textit{monitors}).
    In this example, a citizen (\textit{client}) who wishes to verify the authenticity of this document would engage in an SPMT protocol with the public online record with the information associated with the document (\textit{source-assisted information}) in addition to the document itself.
    This allows the citizen to know that the document is available for public scrutiny without revealing the inspected document.

\subsection{Unsupported use-cases}
\taggedpara{Private DNS resolving}
Private DNS resolving allows a client to query a DNS server without revealing the domain name to the DNS name server.
    This is a use case that is not applicable to our system, as the name server must know the domain name to return the correct IP address.

\taggedpara{Malware detection}
When downloading software from the Internet, verifying that the software is not malicious is crucial.
    Malware detection enables a client to do this by comparing the software's hash with a list of known malicious software hashes.
    However, this also reveals which software the client is downloading, which is undesirable.
    This use case is not applicable to our system, as the server must know the hash of the software to determine that it is not malicious.

%% file: contents/10-conclusion.tex
\section{Conclusion and Future work} \label{sec:conclusion}

This work identifies and examines the SPMT protocol that is relevant in various use cases such as certificate transparency and supply-chain auditing.
By leveraging TEE execution integrity and third-party monitors, we design \sys, a system that does not require the client to disclose their private information to any external parties while achieving constant-time computation and communication.
Our proof-of-concept implementation demonstrates that the proposed system achieves data submission and membership testing latencies of under 7 milliseconds and is capable of serving around 1400 requests per second per CPU core.

Directions for future work include:
(1) Exploring TEE-free design approaches,
(2) Extending the design to support applications that require look-ups, and
(3) Integrating \sys into existing applications.

%% file: contents/12-appendix.tex
\section{Full Correctness Analysis} \label{appendix:correctness}
\subsection{Starting point: Ideal scenario}
We assume an ideal scenario where (1) the communication channel between every entity, (2) the TEE, and (3) the database are ``secure''.
    By ``secure'', we mean the following:
    (1) Secure channel: Total ordering of messages preserved, messages are not delayed or dropped, integrity protected, confidential, and authenticated;
    (2) Secure TEE: No rollback attacks, no forking, no interrupts, no side-channel attacks, and provides confidentiality and execution integrity, remote attestation, state continuity;
    (3) Secure database: No forking and is append-only.

Assume the membership testing process as follows:
    (1) When the data source submits $\dds$ to the secure TEE, the secure TEE sends $\cntpor$ directly to the client via secure channel,
    (2) The secure TEE increments $cnt$ once it finishes storing $\dds$ in the secure database,
    (3) The client requests and receives $\cntpop$ from the secure TEE via secure channel during membership testing, and
    (4) If $\cntpor < \cntpop$ holds, then the algorithm outputs true; otherwise, false.

Such a membership testing process is correct assuming a secure channel, TEE, and database.

\subsection{Removing secure channels (Step 1)}
In this step, we replace a secure channel with digital signatures, losing the total ordering and confidentiality of the messages.

The membership testing process is modified as follows:
    (1) When the data source submits $\dds$ to the secure TEE, the secure TEE creates a digital signature over $\cntpor$, $\dds$, and $report$ (i.e., $POR$),
    (2) The secure TEE increments $cnt$ once it finishes storing $\dds$ in the secure database,
    (3) The client requests and receives a digital signature over $\cntpop$ (i.e., $POP$) from the secure TEE during membership testing, and
    (4) If the digital signatures can be verified using the same TEE public key and $\cntpor < \cntpop$ holds, then the algorithm outputs true; otherwise, false.

Next, we show that the adversary cannot violate the correctness given the new capabilities provided via the removal of secure channels.

\subsubsection{Tampering messages}
The adversary may change the content of the messages, e.g., change the $\cntpor$ value included in $POR$.
    This is easily detectable, as the digital signature cannot be verified using the message.

\subsubsection{Dropping messages}
The adversary may discard messages exchanged between entities.
    Not receiving any information does not affect the correctness of the system.

\subsubsection{Delaying or swapping messages}
This enables the attacker to violate the total order of the exchanged messages.
    For instance, for two $POR$s for two different $\dds$s, $POR^{1}$ and $POR^{2}$, where $\cntpor^{1} < \cntpop < \cntpor^{2}$ holds, the adversary may delay the former with the latter in an attempt to convince that $\cntpor < \cntpop$ holds.
    However, since the $POR$s are created for different $\dds$s, the client will detect this attack by checking whether the $POR$ is for the $\dds$ it is verifying.

\subsubsection{Replaying $POR$}
A malicious data source may provide the client with a $POR$ that was issued with a smaller $\cntpor$ value for a different $\dds$ in an attempt to trick the client into believing that the target $\dds$ has been included in the database when it is not.
    The client can detect this attack, as the provided $POR$ does not match the provided $\dds$.

\subsubsection{Replaying $\{\dds, sig_{\dds}\}$}
This allows an adversary to produce multiple $POR$s with different $\cntpor$ for the same target $\dds$. 
    Assume the adversary generates two $POR$s, $POR^{1}$ and $POR^{2}$: the former with counter value $\cntpor^{1}$ and the latter with $\cntpor^2$ where $\cntpor^1 < \cntpop < \cntpor^2$ holds.
    In an attempt to convince the client that $\cntpor < \cntpop$ holds, they can provide the client $POR^{1}$ instead of $POR^{2}$. 
    This does not violate the correctness of the system, as $\dds$ is in fact stored in the database when the system outputs true.
    
\subsubsection{Replaying $POP$}
An adversary may replace the correct $POP$ with a different one.
    However, since the adversary only has access to $POP$s with $\cntpop$ values \emph{less than or equal} to $\cntpor$, they cannot convince the client that $\cntpor < \cntpop$ is true just by replaying $POP$.

\subsection{Removing secure TEE (Step 2)}
In this step, we replace the secure TEE with a realistic TEE that adheres to the TEE threat model shown in Section~\ref{sec:eval:security:tee-trust}.

The membership testing process is modified as follows:
    (1) When the data source submits $\dds$ to the TEE, the TEE creates a digital signature over $\cntpor$, $\dds$, and $report$ (i.e., $POR$),
    (2) The TEE increments $cnt$ once it finishes storing $\dds$ in the secure database,
    (3) The client requests and receives a digital signature over $\cntpop$ (i.e., $POP$) from the TEE during membership testing, and
    (4) If the digital signatures can be verified using the same TEE public key and $\cntpor < \cntpop$ holds, then the algorithm outputs true; otherwise, false.

Next, we show that the adversary cannot violate the correctness property given the gained new capabilities.

\subsubsection{Leveraging side-channel attacks}
Adversaries may attempt to leverage side-channel attacks to steal secret information from the TEE, enabling the forging of $POR$s or $POP$s.
    This is not possible because we assume that cryptographic operations using TEE signing keys do not leak (Section~\ref{sec:eval:security:tee-trust}).

\subsubsection{Interrupting TEEs}
This enables the malicious server to halt TEEs, preventing $\dds$ from being processed and stored in the database.
    Since halting the TEE halts the progress of $cnt$, $\cntpor < \cntpop$ will never hold, therefore the algorithm will never output \texttt{true}.

\subsubsection{Attacking state-continuity}
This allows an adversary to prevent the TEE state from being passed on after a TEE crash, power-loss, or system-wide reboot.
    Similar to interrupting TEEs, this halts the progress of $cnt$ and, therefore, the client will never receive \texttt{true} from the algorithm.

\subsubsection{Rolling back TEE state}
After receiving the target $\dds$ and producing $POR$, the server may roll the TEE state back before the point it received $\dds$ and continue accepting other data items. We assume a TEE rollback protection mechanism that prevents this (Section~\ref{sec:eval:security:tee-trust}).

\subsubsection{Forking TEEs}
A malicious server may create multiple TEE versions through forking, presenting different versions depending on the interacting party. 
    For example, when the server is about to receive the target data item $\dds$, it can fork the TEE into two versions: $TEE_1$ and $TEE_2$. 
    $TEE_1$ interfaces with the data source to generate a genuine $POR$ for $\dds$, while $TEE_2$ does not receive $\dds$. 
    During membership testing, $TEE_2$ is presented to the client, generating a correct $POP$ that falsely suggests $\cntpor < \cntpop$, even though $\dds$ is not stored in the database. 
    We assume intra-machine and inter-machine fork detection to prevent this (Section~\ref{sec:eval:security:tee-trust}).

\subsubsection{Impersonating TEEs}
This enables adversaries to fake $POR$s or $POP$s or remove $\dds$ from the \texttt{buffer}.
    This attack is prevented because (1) $report$s must be produced by genuine TEEs; (2) $POR$ must include valid $report$s; and (3) $POP$ must be verified using the TEE pubic key included in $report$ sent along with $POR$.

\subsection{Removing secure database (Step 3)}
In this final step, we replace the secure database with a regular database, accumulators, and monitors. %

The membership testing process is modified as follows.
    (1) When the data source submits $\dds$ to the TEE, the TEE creates a digital signature over $\cntpor$, $\dds$, and $report$ (i.e., $POR$),
    (2) The TEE updates the accumulator $HC_T$, stores $HC_T$ and $cnt$ in $List_{HC_T}$, and increments $cnt$ once it finishes storing $\dds$ in an external database,
    (3) The monitor updates its accumulator $HC_M$ with the database content,
    (4) The client requests $HC_M$ from the monitor,
    (5) The client sends $HC_M$ to the TEE, the TEE looks up $\cntpop$ associated with $HC_M$ in $List_{HC_T}$, and sends a digital signature over $\cntpop$ (i.e., $POP$) from the TEE during membership testing, and
    (6) If the digital signatures can be verified using the same TEE public key and $\cntpor < \cntpop$ holds, then the algorithm outputs true; otherwise, false.
This process is equivalent to our solution, except for several primitives that are not related to security, such as the \texttt{buffer} and $ACK$.

Next, we show that the adversary cannot violate the correctness property given the gained capabilities, which concludes our analysis.

\subsubsection{Impersonating monitors}
This enables the attacker to produce an $HC_M$ that satisfies $\cntpor < \cntpop$ although $\dds$ is not in the database.
    This attack is not valid, as we assume the monitor's signing key is protected.

\subsubsection{Dropping \texttt{buffer}, $\dds$ from \texttt{buffer}, or $\dds$ from database} 
An adversary may prevent the \texttt{buffer} that includes a target $\dds$ or only $\dds$ from being sent to the database entirely and send an $ACK$ to the TEE to make it seem that it was sent.
    Since $HC_M$ will not match any $HC_T$ or only match $HC_T$ created \emph{before} \texttt{buffer} was dropped, $\cntpor < \cntpop$ will never hold.
    Furthermore, if clients notice that neither waiting nor contacting other monitors allows them to receive the correct $POP$, they can publicize $\dds$ and $POR$ so that the monitors are aware and can start investigating.

\subsubsection{Forking databases}
This enables an adversary to create multiple versions of the database and show different versions according to whom they are talking to (i.e., split-view attack, see Section~\ref{sec:background:monitors}).
    Since $HC_M$ will not match the most recent $HC_T$ that includes $\dds$, $\cntpor < \cntpop$ will not hold, and therefore the client will not be convinced that $\dds$ is stored in the database.

\subsubsection{Dropping $\dds$ after monitoring}
The adversary may drop the target $\dds$ after the monitor downloads the database content that includes $\dds$.
    While this may cause $HC_M$ to match a $HC_T$ value and cause $\cntpor < \cntpop$ to hold, this is a violation of correctness, since the monitor is aware of $\dds$ and the adversary is caught of its wrongdoings.